\documentclass[
 aps,
 prb,
 amsmath,
 amssymb,
 preprint,
 superscriptaddress
 ]{revtex4-2}

\usepackage[T1]{fontenc}
\usepackage[utf8]{inputenc}
\usepackage[english]{babel}
\usepackage{graphicx}
\usepackage{lmodern}
\usepackage{nicefrac}
\usepackage[version=4]{mhchem}
\usepackage{siunitx}
\usepackage{caption}

\usepackage{makecell}
\usepackage{multirow}

\DeclareSIUnit{\ppm}{ppm}
\DeclareSIUnit{\meV}{meV}
\DeclareSIUnit{\eV}{eV}
\DeclareSIUnit{\Rydberg}{Ry}
\DeclareUnicodeCharacter{2212}{-}
\DeclareUnicodeCharacter{03C0}{$\pi$}
\DeclareUnicodeCharacter{03B1}{alpha}
\DeclareUnicodeCharacter{03B2}{beta}
\DeclareUnicodeCharacter{03B3}{gamma}

\begin{document}
\title{Electronic Structures of Two-Dimensional \ce{PC6}-Type Materials}

\author{Maximilian A. Springer}
\affiliation{Faculty for Chemistry and Food Chemistry, TU Dresden, Bergstrasse 66c, 01069 Dresden, Germany}
\affiliation{Helmholtz-Zentrum Dresden-Rossendorf, Institute of Resource Ecology, Permoserstrasse 15, 04318 Leipzig, Germany}

\author{Thomas Brumme}
\affiliation{Faculty for Chemistry and Food Chemistry, TU Dresden, Bergstrasse 66c, 01069 Dresden, Germany}

\author{Agnieszka Kuc}
\affiliation{Helmholtz-Zentrum Dresden-Rossendorf, Institute of Resource Ecology, Permoserstrasse 15, 04318 Leipzig, Germany}

\author{Thomas Heine}
\email[]{Thomas.Heine@tu-dresden.de}
\affiliation{Faculty for Chemistry and Food Chemistry, TU Dresden, Bergstrasse 66c, 01069 Dresden, Germany}
\affiliation{Helmholtz-Zentrum Dresden-Rossendorf, Institute of Resource Ecology, Permoserstrasse 15, 04318 Leipzig, Germany}
\affiliation{Department of Chemistry, Yonsei University, Seodaemun-gu, Seoul 120-749, Republic of Korea}

\date{\today}

\begin{abstract}
Two-dimensional (2D) materials may exhibit intriguing band structure features (\textit{e.g.}, Dirac points), that lay far away from the Fermi level. They are, thus, not usable in applications. The semiconducting 2D material \ce{PC6} has two Dirac cones above and below the Fermi level. Therefore, it is an ideal playground to demonstrate chemical functionalization methods for shifting the Fermi level in order to access interesting band structure features. \ce{PC6} is based on the $\sqrt{7}\times\sqrt{7}R19.1^\circ$ super cell of graphene with two carbon atoms per unit cell substituted by phosphorous. It is demonstrated how substitution with other heteroatoms that contain a different number of valence electrons, the Dirac points can be accessed. Alternatively, hydrogen atoms can be used as adatoms at the heteroatom sites. This increases electron filling and shifts the Fermi level upwards.
\end{abstract}

\maketitle

\section{Introduction}
Graphene monolayer exhibits intriguing electron transport phenomena, \textit{e.g.}, massless fermi\-ons at the K point~\cite{novoselov_two-dimensional_2005}. The associated band structure feature is the Dirac cone~\cite{wehling_dirac_2014}. This leads to high carrier mobility for both electrons and holes and the presence of Quantum Hall effect at room temperature in strong magnetic fields~\cite{zhang_experimental_2005,novoselov_unconventional_2006,novoselov_room-temperature_2007}. However, the band gap, which is induced by spin-orbit coupling (SOC), is very small~\cite{kane_quantum_2005}. Therefore, pristine graphene is not very useful in applications like transistors or switching devices. To make it helpful for such applications, the band gap has to be increased. Hence, there are significant research efforts to tune the band structure of graphene~\cite{liu_chemical_2011,novoselov_roadmap_2012}. Methods to open the band gap, \textit{e.g.}, chemical functionalization, received a lot of attention~\cite{georgakilas_functionalization_2012}. For example, the $sp^2$ hybridization of carbon atoms in graphene can be exploited by attaching other atoms or molecules. The addition of hydrogen or fluorine to all carbon atoms gives rise to graphane and fluorographane, respectively~\cite{sofo_graphane_2007,robinson_properties_2010}. In graphane, all carbon atoms are $sp^3$ hybridized, which means that the formerly planar structure is puckered, with carbon atoms above and below the plane. Graphane is an insulator with a band gap of $5.4\,\si{\electronvolt}$~\cite{lebegue_accurate_2009}. Additionally, graphene layers can be stacked. Bilayers and few-layer stacks got substantial attention for their potential use as supercapacitors and superconductors~\cite{wang_three_2013, cao_unconventional_2018}. Furthermore, structural modifications of the lattice itself, \textit{e.g.}, antidot lattices, were investigated~\cite{furst_electronic_2009}.

The intriguing charge transport properties of graphene have lead to further research in the field of two-dimensional (2D) materials beyond graphene~\cite{butler_progress_2013,chhowalla_chemistry_2013}. 2D polymers, \textit{e.g.}, the 2D triphenyl-bismuth lattice~\cite{wang_organic_2013} can exhibit Dirac points as well~\cite{springer_topological_2020}. Here, the Dirac cone is gapped by $43\,\si{meV}$ and lies $0.31\,\si{eV}$ below the Fermi level ($E_\mathrm{F}$). Similarly, the \ce{Ni3C12S12} kagome lattice has a Dirac cone about $0.5\,\si{eV}$ above the Fermi level~\cite{wang_prediction_2013}. This system is especially interesting, since SOC induces a topological band gap opening at the Dirac point. That means the \ce{Ni3C12S12} lattice is an organic topological insulator. However, the band gap features are only usable if they lie in the vicinity of the Fermi level, which is not the case in the examples of the triphenyl-bismuth lattice and \ce{Ni3C12S12}. While graphene analogs and materials with intriguing properties are proposed, they might bear the problem that intriguing phenomena lie far away from the Fermi level. Therefore, strategies to shift the Fermi level are of central importance for the field. Yet, there are five popular strategies to shift the Fermi level: if the feature of interest is close enough to standard electron filling, strain can slightly change the band structure. Thus, Dirac points can be shifted towards the Fermi level~\cite{wang_tungsten_2019}. Alternatively, electrons can be added or deducted by gate voltages and electric fields as well~\cite{ueno_electric-field-induced_2008, avetisyan_electric-field_2009, yu_tuning_2009, goldman_electrostatic_2014}. Besides these physical interventions, chemical modifications are also possible. These are, for example, the inclusion of adsorbates~\cite{leenaerts_adsorption_2008, lherbier_charge_2008, wehling_molecular_2008, gierz_atomic_2008, park_single-gate_2012, lu_semiconducting_2013}, building interfaces with surfaces leading to charge separation~\cite{chen_surface_2007}, or doping~\cite{wei_synthesis_2009}.

Recently, the 2D phosphorous carbide \ce{PC6} received significant atten\-tion~\cite{yu_two-dimensional_2019}. Once alkali metal ions, \textit{e.g.}, potassium or lithium, are adsorbed on \ce{PC6}, the semiconducting material becomes metallic~\cite{dou_prediction_2019}. Furthermore, it exhibits small ion diffusion barriers. Therefore, it is discussed as anode material in lithium ion batteries~\cite{dou_prediction_2019, fan_monolayer_2020, zhang_two-dimensional_2020}. Additionally, \ce{PC6} can act as a gas sensor due to the  conductivity being highly sensitive towards the addition of adsorbents~\cite{yu_pc6_2020, zhou_carbon_2020}. However, intrinsic properties of the material's network structure where not yet discussed. It can be considered as functionalized graphene, since its structure can be derived from a graphene super cell with two carbon atoms per cell substituted by heteroatoms. Due to the retained honeycomb topology, it has Dirac points above and below $E_\mathrm{F}$. Therefore, it can be used as a playground to investigate strategies for shifting the Fermi level towards these Dirac points: the general shape of the band structure is retained if substitutions are isostructural. Therefore, $E_\mathrm{F}$ can be controlled by using other heteroatoms than phosphorous, which change the electron filling due to their different number of valence electrons. Alternatively, hydrogen adatoms can be used to add valence electrons.

\section{Theoretical Methods}
First-principles calculations within the density-functional theory (DFT) framework were employed for both full structural optimization (atomic positions and lattice vectors) and electronic properties using AMS/BAND 2019~\cite{te_velde_precise_1991,philipsen_band2019_nodate, kadantsev_formulation_2007}. For optimizations, the Perdew-Burke-Ernzer\-hof (PBE) functional was used with the TZP Slater-type basis set and Grimme-D3 dispersion energy correction with Becke-Johnson damping~\cite{perdew_generalized_1996, perdew_restoring_2008, grimme_effect_2011, franchini_becke_2013, franchini_accurate_2014}. The necessary accuracy of the tetrahedron method based \textit{k}-space grid was determined for each system separately to a precision of \num{e-3}~\si{\eV}. In order to account for relativistic effects in systems with gallium, germanium, arsenic and selenium, the scalar zeroth-order regular approximation (ZORA) was used in relaxation calculations with AMS/BAND~\cite{philipsen_relativistic_1997, philipsen_relativistic_2000}. For electronic structure calculations, the Heyd-Scuseria-Ernzerhof functional (HSE06) was used~\cite{heyd_hybrid_2003,heyd_erratum_2006}. For phonon calculations, both lattice and atomic positions were relaxed using the PBE functional and the projector-augmented wave (PAW) method as implementend in QuantumEspresso 6.6 with a $5 \times 5 \times 1$ \textit{k}-grid, $70\,\si{\Rydberg}$ wave function cutoff and $580\,\si{\Rydberg}$ kinetic energy cutoff~\cite{giannozzi_advanced_2017}. Phonon band structures were obtained from phonopy using the finite displacement method~\cite{togo_first_2015}. In order to converge the phonon band structures, forces of the $5 \times 5 \times 1$ super cells were evaluated based on the $\Gamma$-point approximation.

\section{Results and Discussion}
\subsection{Structure of \ce{PC6}}
\begin{figure}
  \centering
    \includegraphics[width=0.7\textwidth]{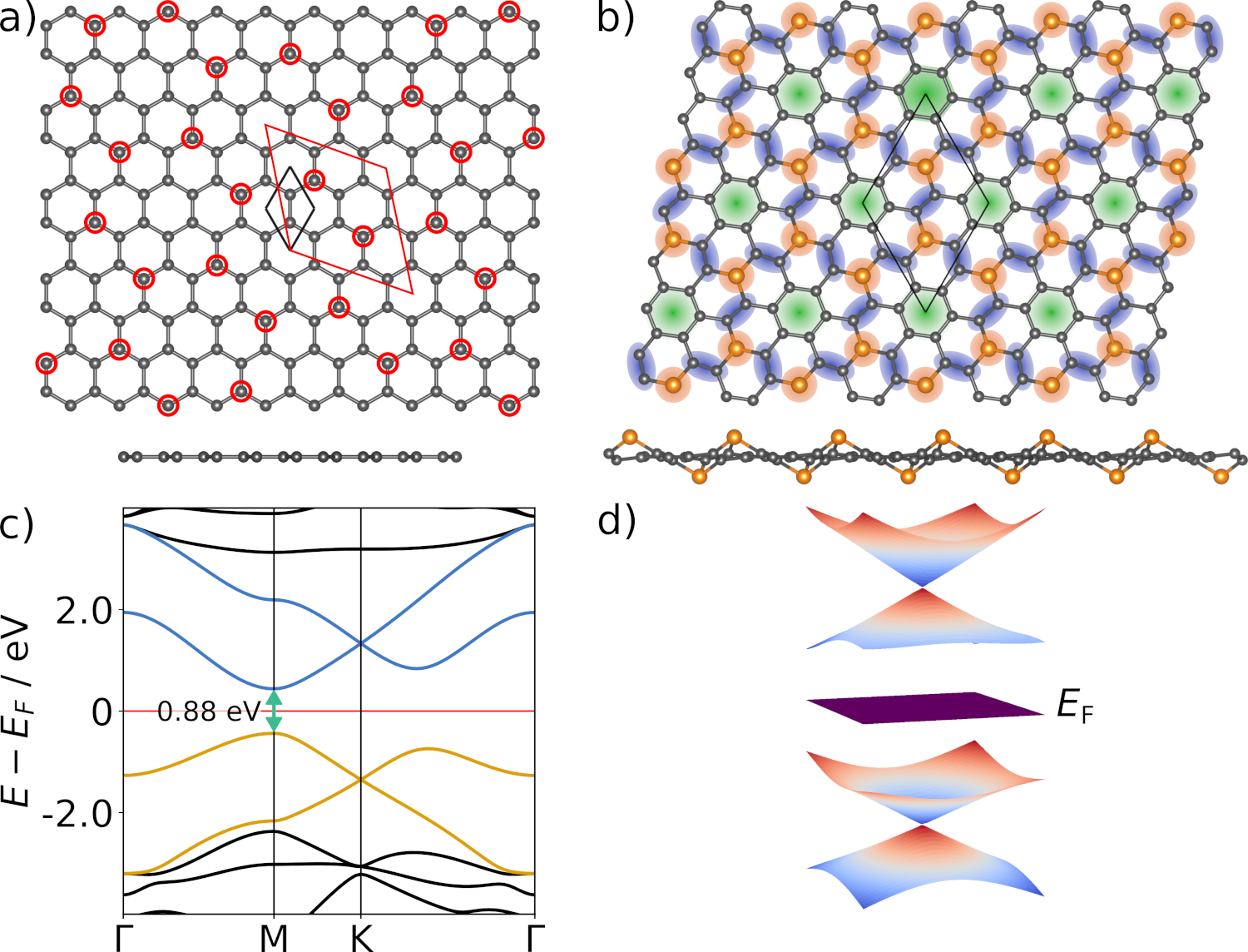}
    \caption{(a) Top and side view of graphene with the primitive unit cell (black lines) and  the $\sqrt{7}\times\sqrt{7}R19.1^\circ$ super cell (red lines). Carbon atoms forming a honeycomb super lattice are marked by red circles. (b) Top and side view of chemical structure of \ce{PC6} with the hexagons formed by phosphorous (orange background color), the kagome lattice formed by carbons next to phosphorous (blue) and the carbon atoms forming the hexagonally arranged six-membered rings (green), the side view reveals the puckered structure with phosphorous atoms above and below the plane of carbon atoms. (c) Band structure of \ce{PC6} with direct band gap at M point ($E_\mathrm{g} = 0.88\,\si{eV}$) and the characteristic bands forming the Dirac cone above (blue bands) and below (golden bands) the Fermi level. (d) 3D band structure of \ce{PC6} at K point, showing the four bands above and below the Fermi level ($E_\mathrm{F}$, purple plane). The 3D band structure was calculated with AMS/DFTB~\cite{ruger_ams_nodate} with 3ob-3-1 parameters~\cite{gaus_dftb3_2011}. Carbon atoms are grey, phosphorous atoms are orange.}
    \label{fig:PC6_fig1}
\end{figure}

In Fig.~\ref{fig:PC6_fig1}a, the $\sqrt{7}\times\sqrt{7}R19.1^\circ$ super cell of graphene is highlighted by red lines. Based on this super cell, a second honeycomb network can be formed. If two out of the 14 carbon atoms per cell (red circles in Fig.~\ref{fig:PC6_fig1}a) are substituted by, \textit{e.g.}, phosphorous, a network with three sublattices is formed (\textit{cf.}~Fig.~\ref{fig:PC6_fig1}b): a honeycomb super lattice of phosphorous atoms (orange background), a kagome lattice formed by carbon atoms connecting the phosphorous atoms to each other (blue) and hexagonal lattice of carbon atoms forming six-membered rings (green). Due to lone electron pairs at the phosphorous atom, the structure is puckered with phosphorous atoms above and below the plane formed by carbon atoms (\textit{cf.}~side view in Fig.~\ref{fig:PC6_fig1}b). In Table~\ref{tab:PC6_plain_structures}, structural parameters and band gap of \ce{PC6} are displayed with data for other \ce{PC6}-type systems. The lattice parameter of \ce{PC6} is $6.69\,\si{\angstrom}$, which is slightly larger than the parameter of the super cell of graphene ($6.51\,\si{\angstrom}$). This is due to a small increase in the length of C-C bonds ($1.47\,\si{\angstrom}$ in \ce{PC6} vs.~$1.42\,\si{\angstrom}$ in graphene). The structure is puckered, which means that phosphorous atoms have an offset ($\Delta z$) of $1.09\,\si{\angstrom}$ from the plane of carbon atoms. The C-P bonds are $1.82\,\si{\angstrom}$ long, which is similar to other phosphorous carbides, for which bond lengths between $1.76$ and $1.86\,\si{\angstrom}$ were reported from computational studies~\cite{guan_two-dimensional_2016}. Hirshfeld charges are $q=0.19$ for the phosphorous atoms, $q=-0.06$ for carbon atoms on the kagome sublattice and almost zero for atoms on the hexagonal sublattice (\textit{cf.}~Table \ref{tab:SI_PC6_plain_charges}). Despite its puckered structure, \ce{PC6} is still $\pi$ conjugated. It is a semiconductor with direct band gap at M point ($E_\mathrm{g} = 0.88\,\si{\electronvolt}$). It has two gapped Dirac points above (formed by the blue bands in Fig.~\ref{fig:PC6_fig1}c, $1.33\,\si{\electronvolt}$ above $E_\mathrm{F}$) and below the Fermi level (formed by golden bands in Fig.~\ref{fig:PC6_fig1}c, $1.35\,\si{\electronvolt}$ below $E_\mathrm{F}$). The effective masses for electrons and holes are in the range of $\nicefrac{m^*}{m_\mathrm{e}} = 0.2$-0.5 with the electron rest mass $m_\mathrm{e}$. In Fig.~\ref{fig:PC6_fig1}d, a three-dimensional (3D) band structure is shown with the Fermi level indicated by the purple plane. The 3D band structure confirms that the crossings above and below the Fermi level are indeed Dirac points. Herein, we present two chemical means to control the Fermi level in \ce{PC6}-based systems: it can be shifted by occupying the honeycomb super lattice with other chemical species or by adsorption.

\begin{table}[]
    \centering
    \caption{Lattice constant $a$, offset of heteroatom w.r.t carbon plane $\Delta z$,  bond length between heteroatoms and neighboring carbon atoms $d(\mathrm{C-X})$, and band gap $E_\mathrm{g}$ for \ce{PC6}-type 2D materials.}\label{tab:PC6_plain_structures}
    \begin{tabular}{|c|c|c|c|c|}\hline
    	&	$a$ / $\si{\angstrom}$	&	$\Delta z$ / $\si{\angstrom}$	&	$d(\mathrm{C-X})$ / $\si{\angstrom}$ & $E_\mathrm{g}$ / $\si{\electronvolt}$\\\hline
\ce{AlC6}	&	6.877	&	1.032	&	1.922 & 1.95\\\hline
\ce{SiC6}	&	7.013	&	0.000	&	1.732 & none\\\hline
\ce{PC6}	&	6.687	&	1.090	&	1.819 & 0.88\\\hline
\ce{SC6}	&	6.699	&	0.931	&	1.783 & none\\\hline
\ce{GaC6}	&	6.849	&	1.138	&	1.965 & 1.72\\\hline
\ce{GeC6}	&	6.757	&	1.280	&	1.953& none\\\hline
\ce{AsC6}	&	6.711	&	1.340	&	1.974 & 1.33\\\hline
\ce{SeC6}	&	6.778	&	1.158	&	1.940& none\\\hline
    \end{tabular}
\end{table}

\subsection{Substitution Strategy}
The bands around the Fermi level are almost purely formed by $p_z$ orbitals with only small contributions from other orbitals, as reported by Yu \textit{et al.}~\cite{yu_two-dimensional_2019} Therefore, the shape of the band structure does not depend on the exact chemical composition, but on the network topology and the retention of the $\pi$ conjugation~\cite{springer_topological_2020}. The Fermi level can, thus, be shifted by substituting phosphorous atoms in \ce{PC6} with other tetrahedrally (resulting in puckered structures) or trigonally (resulting in planar structures) coordinated heteroatoms.

According to the number of valence electrons relative to phosphorous, the Fermi level can be shifted upwards or downwards compared to the electron filling in \ce{PC6}. In order to demonstrate this effect, compounds of the \ce{PC6}-type with heteroatoms from group 14 (\ce{AlC6}, \ce{GaC6}), group 15 (\ce{SiC6}, \ce{GeC6})~\cite{yang_g-sic6_2021}, group 16 (\ce{PC6}, \ce{AsC6}), and group 17 (\ce{SC6}, \ce{SeC6}) are presented (structures and band structures are shown in Fig.~\ref{fig:SI_PC6_plains} in the Supporting Information). Structural parameters are shown in Table \ref{tab:PC6_plain_structures}. For the considered systems, the cell parameters are in the range $6.70-7.01\,\si{\angstrom}$. As in \ce{PC6}, this is larger than the parameter for the $\sqrt{7}\times\sqrt{7}R19.1^\circ$ super structure of graphene. The lowest value is found for the system with the heteroatom of smallest ionic radius, \ce{SC6}. However, the largest value is found for \ce{SiC6}, the only flat system. In all other structures, the heteroatoms are displaced with respect to the plane formed by carbon atoms. Here, the heteroatoms are $0.93\,\si{\angstrom}$ (\ce{SC6}) to $1.34\,\si{\angstrom}$ (\ce{AsC6}) above and below the plane, respectively. Similarly, bond lengths between heteroatoms and carbon atoms range between $1.73\,\si{\angstrom}$ (\ce{SiC6}) and $1.97\,\si{\angstrom}$ (\ce{AsC6}). All discussed structural parameters loosely follow the trend of atomic radii. Cell parameter, displacement from the plane, and bond length scale with the steric demand, and, thus, the atomic radius of the respective heteratom. The atomic radii decrease from smaller to larger number of valence electrons in each row and increase for ascending number of electrons in each group of the periodic table of elements.

The charge distribution can be investigated with Hirshfeld charges. They show that all systems are carbides with positively charged heteroatoms and negatively charged carbon atoms (\textit{cf.}~Table \ref{tab:SI_PC6_plain_charges} in the Supporting Information).

\begin{figure}
    \centering
    \includegraphics[width=0.7\textwidth]{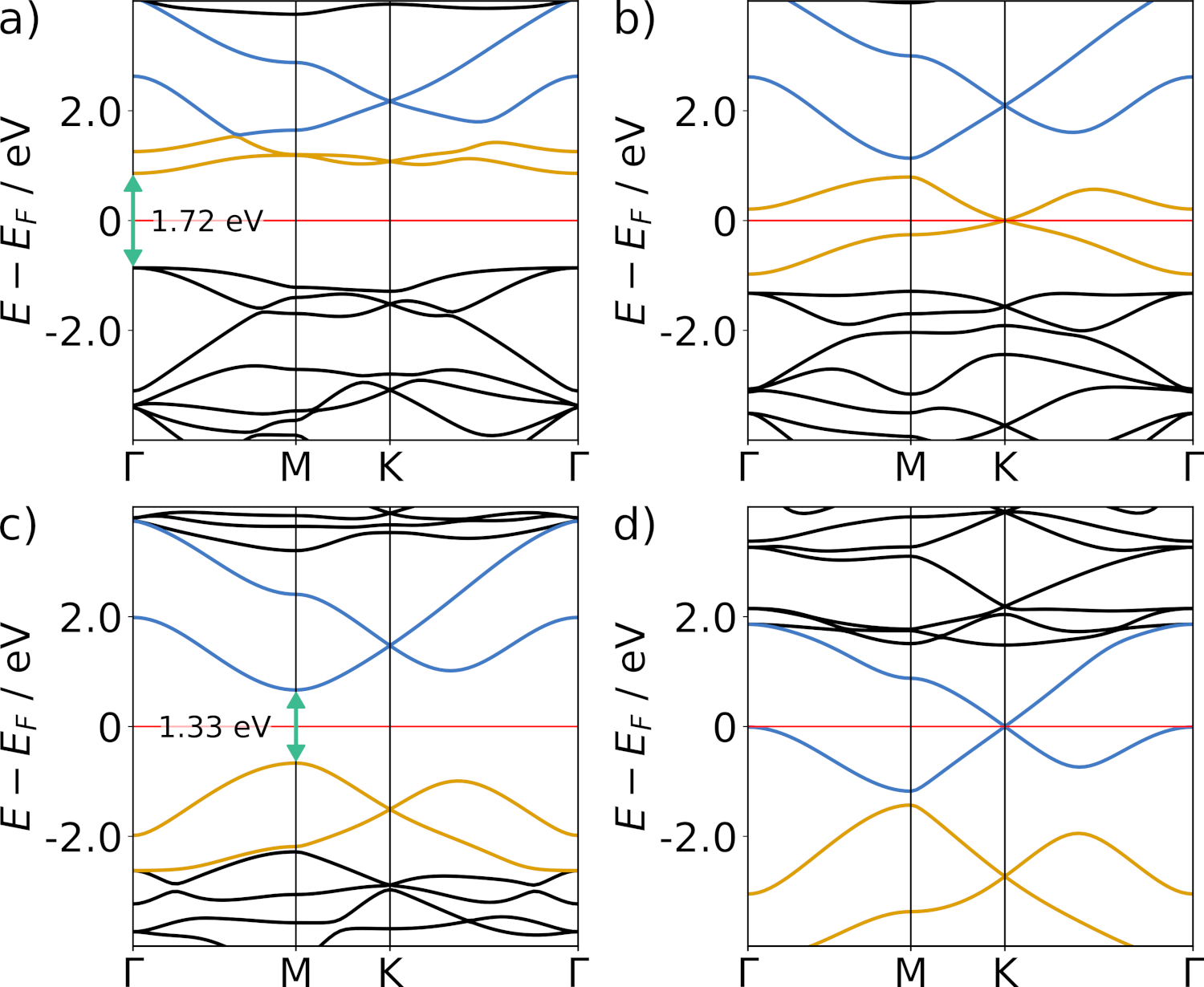}
    \caption{Band structures of (a) \ce{GaC6}, (b) \ce{GeC6}, (c) \ce{AsC6}, and (d) \ce{SeC6}. The characteristic bands are highlighted as in Fig.~\ref{fig:PC6_fig1}c. Increasing the number of valence electrons shifts the Fermi level upwards. Each step corresponds to two additional valence electrons. Thus, Dirac materials are obtained for \ce{GeC6} and \ce{SeC6}.}
    \label{fig:PC6_fig2}
\end{figure}

In Fig.~\ref{fig:PC6_fig2}, band structures for the fourth row carbides \ce{GaC6}, \ce{GeC6}, \ce{AsC6}, and \ce{SeC6} are shown (in the order of ascending number of electrons). The four characteristic bands from the \ce{PC6} band structure (blue and golden bands in Fig.~\ref{fig:PC6_fig1}) can be found in these band structures, as well. As expected, the Fermi level can be shifted upwards or downwards depending on the number of valence electrons: \ce{GaC6} (Fig.~\ref{fig:PC6_fig2}a) has four valence electrons per unit cell less than \ce{PC6}, which means that the Fermi level lies below the four characteristic bands. It is a direct band gap semiconductor ($\Gamma \rightarrow \Gamma$) with $E_\mathrm{g} = 1.72\,\si{\electronvolt}$. For \ce{GeC6}, which has two valence electrons per unit cell less than \ce{PC6}, a metallic system is found. The Fermi level cuts the Dirac cone of the two bands which were below $E_\mathrm{F}$ in \ce{PC6} (golden bands in Fig.~\ref{fig:PC6_fig2}b). Concerning the valence electrons, \ce{AsC6} is isoelectronic to \ce{PC6}. Therefore, it is a semiconductor with direct band gap of $1.33\,\si{\electronvolt}$ (M $\rightarrow$ M, \textit{cf.}~Fig.~\ref{fig:PC6_fig2}c). Following that principle, \ce{SeC6}, which has two valence electrons more than \ce{PC6} and \ce{AsC6}, is a Dirac material with the Fermi level at the upper Dirac cone (golden bands above the Fermi level in \ce{PC6}, \textit{cf.}~Fig.~\ref{fig:PC6_fig2}d).

For semiconducting systems, effective masses were calculated as an indicator of electron transport properties (\textit{cf.}~Table \ref{tab:SI_PC6_plain_effective}). For the two investigated group 14 carbides, \ce{AlC6} and \ce{GaC6}, hole effective masses are $\nicefrac{m^*_\mathrm{h}}{m_\mathrm{e}} = 6.7$. The electron effective masses are $\nicefrac{m^*_\mathrm{e}}{m_\mathrm{e}} = 1.1$ (\ce{AlC6}) and $\nicefrac{m^*_\mathrm{e}}{m_\mathrm{e}} = 1.0$ (\ce{GaC6}). Effective masses determined for group 16 carbides (\ce{PC6} and \ce{AsC6}) are much smaller. In both cases, hole effective masses at the $\Gamma \rightarrow$ M edge are around $\nicefrac{m^*_\mathrm{h}}{m_\mathrm{e}} = 0.5$ and at the $\Gamma \rightarrow$ K edge, they are around $\nicefrac{m^*_h}{m_\mathrm{e}} = 0.2$. Electron effective masses are similarly small.

Transport properties of Dirac materials can be characterized by the Fermi velocity. It is calculated from the slope of linear bands using the formula $v_\mathrm{F} = \frac{1}{\hbar}\frac{\partial E}{\partial k}$. In order to facilitate the discussion, average values for $k_x$ and $k_y$ directions are given. Found Fermi velocities are all well above $10^6\,\si{\meter\per\second}$, with the lowest one found for \ce{SiC6} ($v_\mathrm{F} = 2.7 \times 10^6\,\si{\meter\per\second}$) and the largest found for \ce{GeC6} ($v_\mathrm{F} = 9.9 \times 10^6\,\si{\meter\per\second}$). The velocities for \ce{SC6} and \ce{SeC6} are between those two values, determined to be $3.8 \times 10^6\,\si{\meter\per\second}$ and $3.9 \times 10^6\,\si{\meter\per\second}$, respectively.

\ce{NC6} and \ce{SiC6} are exceptions to the systematics of shifting the Fermi level by isostructural substitution. These structures are flat instead of puckered (\textit{cf.}~Fig.~\ref{fig:SI_PC6_flat} in the Supporting Information). That means the heteroatoms lie in the plane of carbon atoms. For \ce{NC6}, this results in a metallic system. Bands separated by the global band gap in \ce{PC6} overlap energetically in \ce{NC6}. For \ce{SiC6}, however, the Fermi level seems to be shifted upwards instead of downwards according to the lower number of valence electrons compared to \ce{PC6}. Due to the offset from the carbon plane, the overlap of $p_z$ atomic orbitals is reduced compared to a flat structure. This distortion can change the electronic structure. For an artificially puckered \ce{SiC6} structure, the expected band structure was obtained with the Fermi level crossing the lower Dirac point (w.r.t band structure of \ce{PC6}, \textit{cf.}~Fig.~\ref{fig:PC6_fig1}).

\begin{figure}
    \centering
    \includegraphics[width=0.7\textwidth]{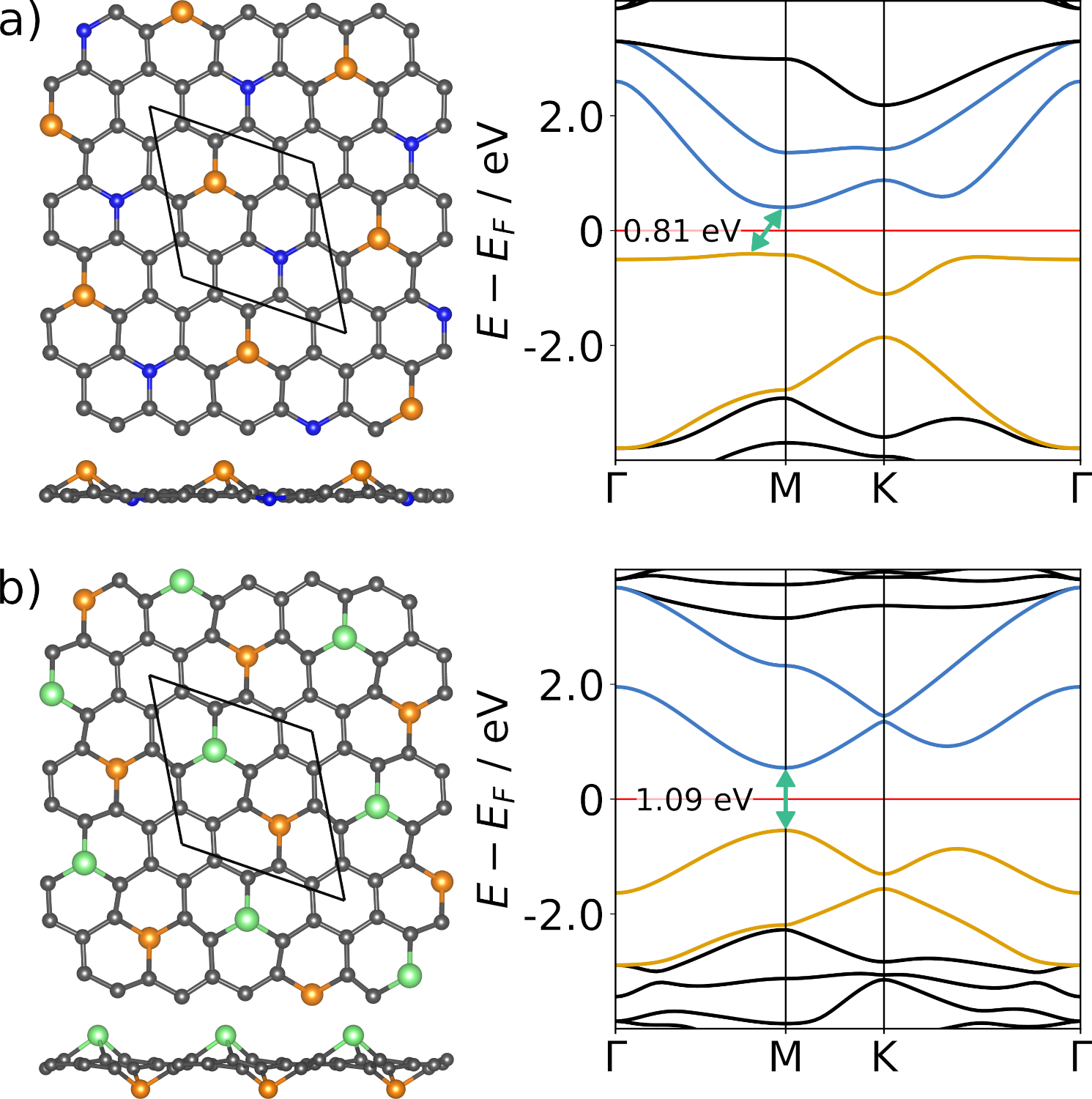}
    \caption{Structures (left) and band structures (right) for carbides with two types of heteratoms, (a) \ce{PNC12} and (b) \ce{PAsC12}. Both systems are semiconductors with the Fermi level between the characteristic bands (coloring according to Fig.~\ref{fig:PC6_fig1}c).}
    \label{fig:PC6_mixed}
\end{figure}

While the curvature of bands can be influenced by substituting only one phosphorous atom per unit cell with nitrogen or arsenic, the position of the Fermi level cannot be influenced. In Fig.~\ref{fig:PC6_mixed}, the structure and band structure of \ce{PNC12} and \ce{PAsC12} are shown, respectively. Lattice parameters and P-C bond lengths are similar to the ones in \ce{PC6}. The N-C and As-C bond lengths are $1.4\,\si{\angstrom}$ in \ce{PNC12} and $2.0\,\si{\angstrom}$ in \ce{PAsC12}, respectively. This is comparable to the known carbon nitride modification \ce{g-C3N4} and arsenic carbides, respectively (\textit{cf.}~Table \ref{tab:PC6_mixed_structures})~\cite{teter_low-compressibility_1996, charifi_theoretical_2009}. Likewise, the offset from the carbon plane is much smaller for the nitrogen atom in \ce{PNC12} ($0.24\,\si{\angstrom}$) than for the arsenic atom in \ce{PAsC12} ($1.29\,\si{\angstrom}$). For both systems, the phosphorous atoms are positively and carbon atoms are mainly negatively charged. However, in \ce{PNC12}, nitrogen atoms carry almost no charge (\textit{cf.}~Fig.~\ref{fig:SI_PC6_mixed_charges}].

\begin{table}[]
    \centering
    \caption{Lattice constant $a$, offset of phosphorous atom w.r.t the carbon plane $\Delta z_\mathrm{P}$, offset of heteroatom w.r.t the carbon plane $\Delta z_\mathrm{X}$, bond length between phosphorous and neighboring carbon atoms $d(\mathrm{C-P})$, and bond length between heteroatom and neighboring carbon atoms $d(\mathrm{C-X})$.}
    \label{tab:PC6_mixed_structures}
    \begin{tabular}{|c|c|c|c|c|c|c|}\hline
    	&	$a$ / $\si{\angstrom}$	&	$\Delta z_\mathrm{P}$ / $\si{\angstrom}$	& $\Delta z_\mathrm{X}$ / $\si{\angstrom}$	&	$d(\mathrm{C-P})$ / $\si{\angstrom}$ & $d(\mathrm{C-X})$ / $\si{\angstrom}$\\\hline
\ce{PNC12} & 6.594 & 0.939 & 0.244 & 1.789 & 1.418\\\hline
\ce{PAsC12} & 6.697 & 1.147 & 1.290 & 1.834 & 1.957\\\hline
    \end{tabular}
\end{table}

Both \ce{PNC12} and \ce{PAsC12} are semiconductors with band gaps of $0.81\,\si{\electronvolt}$ and $1.09\,\si{\electronvolt}$, respectively. While both conduction band minimums are at the M point, the valence band maximum in \ce{PNC12} lies on the edge between $\Gamma$ and M point. In this system, the hole effective masses are above $1$ and the electron effective masses are below 1 (\textit{cf.}~Table \ref{tab:SI_PC6_mixed_electronics}). \ce{PAsC12} is a direct band gap semiconductor (M $\rightarrow$ M) with effective masses as found for \ce{PC6}.

\subsection{Adatom Strategy}
\begin{table}[]
    \centering
    \caption{Lattice constant $a$, offset of heteroatom w.r.t carbon plane $\Delta z_\mathrm{X}$, bond length between heteroatoms and hydrogen adatoms $d(\mathrm{X-H})$.}\label{tab:PC6_plain_adatom_structures}
    \begin{tabular}{|c|c|c|c|c|}\hline
    	&	$a$ / $\si{\angstrom}$	&	$\Delta z_\mathrm{X}$ / $\si{\angstrom}$	&	$d(\mathrm{X-H})$ / $\si{\angstrom}$\\\hline
\ce{HPC6}	&	6.799	&	0.754	&	1.443\\\hline
\ce{HAsC6}	&	6.859	&	0.966	&	1.550\\\hline
    \end{tabular}
\end{table}

\begin{figure}
    \centering
    \includegraphics[width=0.7\textwidth]{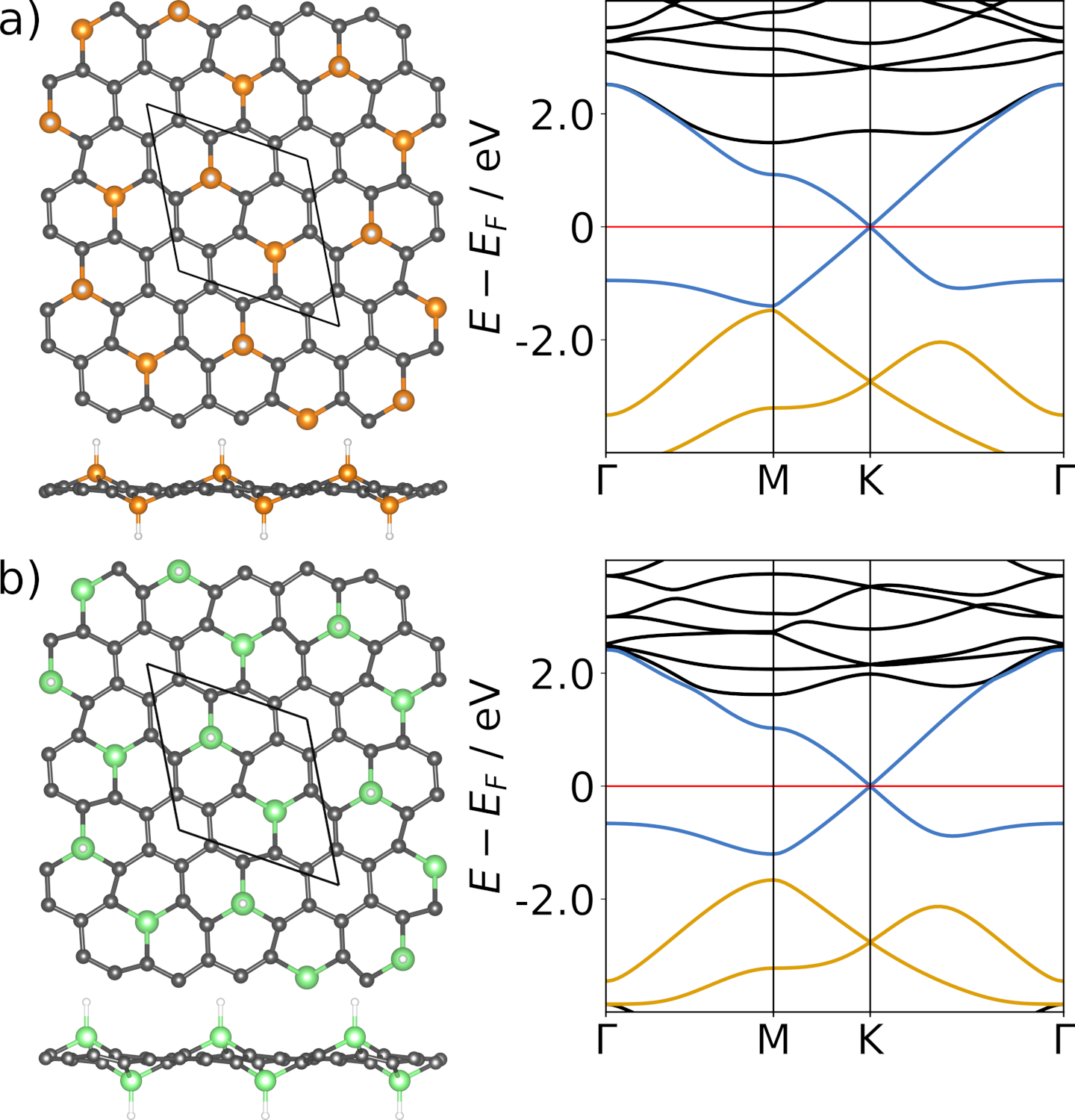}
    \caption{Structures (left) and band structures (right) of (a) \ce{HPC6} and (b) \ce{HAsC6}. Addition of hydrogen increases the number of valence electrons per unit cell by two, which shifts the Fermi level one band upwards compared to the electron filling in \ce{PC6} and \ce{AsC6}, respectively. Coloring of bands according to Fig.~\ref{fig:PC6_fig1}b. Hydrogen is white, carbon is grey, phosphorous is orange, and arsenic is green.}
    \label{fig:PC6_plain_adatoms}
\end{figure}

Alternative to the substitution strategy, heteroatoms in the group 15 carbides (phosphorous and arsenic) can be used as binding sites for hydrogen atoms~\cite{lu_hydrogenated_2021}. Structural parameters of \ce{HPC6} and \ce{HAsC6} are shown in Table \ref{tab:PC6_plain_adatom_structures}. For both systems, the lattice parameter is slightly larger than in the pristine structures. This is also reflected in the smaller offset of heteroatoms from the carbon plane, which is reduced from $1.09\,\si{\angstrom}$ in \ce{PC6} to $0.75\,\si{\angstrom}$ in \ce{HPC6} and from $1.34\,\si{\angstrom}$ in \ce{AsC6} to $0.97\,\si{\angstrom}$ in \ce{HAsC6}. The distance between heteroatom and hydrogen atom is in the range of $1.4$ to $1.6\,\si{\angstrom}$, which is similar to the bond lengths in phosphine and arsenine, respectively (\textit{cf.}~Fig.~\ref{fig:SI_PC6_phosphine-arsenine} in the Supporting Information). Hirshfeld charges for the hydrogen atoms are almost zero (\textit{cf.}~Fig.~\ref{fig:SI_PC6_hydrogenated_charges}). The two additional hydrogen atoms increase the number of electrons in the unit cell, while network structure and symmetry remain intact. As it can be seen in the band structures for both \ce{HPC6} and \ce{HAsC6} in Fig.~\ref{fig:PC6_plain_adatoms}, the general shape of the band structure is retained and the Fermi level is shifted to the upper Dirac cone. The Fermi velocity for both systems is $v_\mathrm{F} = 3.2 \times 10^6\,\si{\meter\per\second}$.

\begin{figure*}
    \centering
    \includegraphics[width=0.7\textwidth]{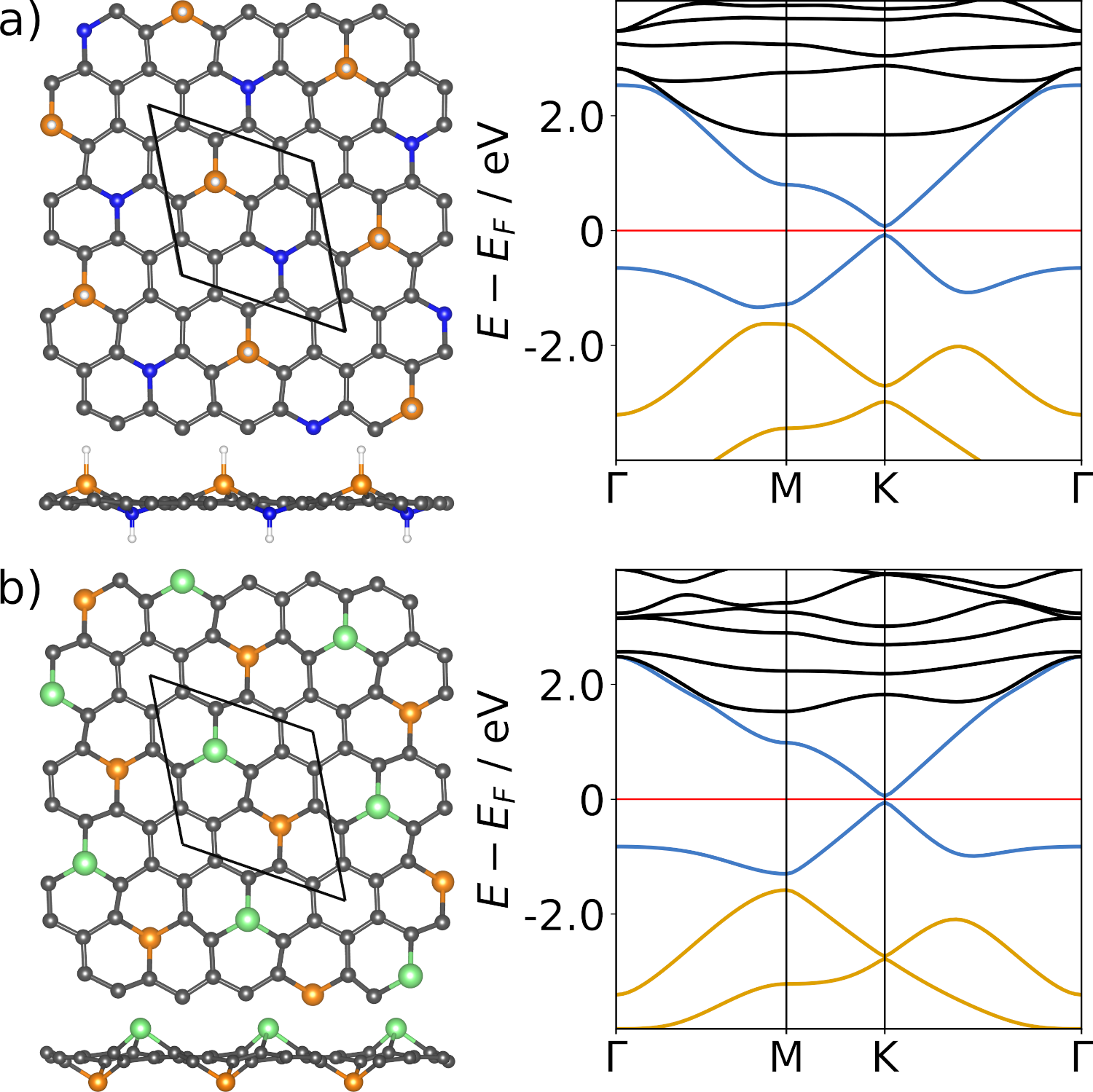}
    \caption{Structures (left) and band structures (right) of (a) \ce{H2PNC12} and (b) \ce{H2PAsC12}. Addition of hydrogen increases the number of valence electrons per unit cell by 2, which shifts the Fermi level one band upwards compared to the electron filling in \ce{PNC12} and \ce{PAsC12}, respectively. Due to the broken inversion symmetry, the Dirac points are slighly gapped (\ce{H2PNC12}: $E_\mathrm{g} = 0.15\,\si{\electronvolt}$, \ce{H2PAsC12}: $E_\mathrm{g} = 0.11\,\si{\electronvolt}$). Coloring of bands according to Fig.~\ref{fig:PC6_fig1}b. Hydrogen is white, carbon is grey, nitrogen is blue, phosphorous is orange, and arsenic is green.}
    \label{fig:PC6_mixed_adatoms}
\end{figure*}

Analogously, hydrogen atoms can be added to the unsymmetrically substituted systems, which yields \ce{H2PNC12} and \ce{H2PAsC12} (\textit{cf.}~Fig.~\ref{fig:PC6_mixed_adatoms}). The discussed structural properties follow the same trends as in \ce{HPC6} and \ce{HAsC6}. In the respective band structures for \ce{H2PNC12} and \ce{H2PAsC12} (\textit{cf.}~Fig.~\ref{fig:PC6_mixed_adatoms}), the Fermi level is also shifted upwards with respect to the band structures of the pristine systems. However, the Dirac points are gapped (\ce{H2PNC12}: $E_\mathrm{g} = 0.15\,\si{\electronvolt}$, \ce{H2PAsC12}: $E_\mathrm{g} = 0.11\,\si{\electronvolt}$). The existence of a Dirac cone is closely related to inversion symmetry and time-reversal symmetry. Inversion symmetry is broken due to the unsymmetric substitution. As remnants from Dirac points, the effective masses are much smaller than for the pristine 2D materials, being as low as $m^* = 0.02$ in \ce{H2PAsC12} (\textit{cf.}~Table \ref{tab:SI_PC6_hyd_mixed_effective}).

\subsection{Topological Properties}
Many of the presented systems are Dirac materials. Therefore, there was the possibility of obtaining topological quantum materials. Hence, an analysis of the wave function at the TRIM points was conducted~\cite{fu_topological_2007}. However, none of the Dirac materials \ce{SiC6}, \ce{SC6}, \ce{GeC6}, \ce{SeC6}, \ce{HPC6}, or \ce{HAsC6} turned out to have a non-zero $Z_2$ invariant. Therefore, the systems are topologically trivial.

\section{Conclusions}
 
By substitution with heteroatoms, a new family of 2D materials can be derived from the $\sqrt{7}\times\sqrt{7}R19.1^\circ$ super cell of graphene. Due to the substitution of carbon atoms with heteroatoms, \textit{e.g.}, phosphorous, a honeycomb super structure arises in a $\pi$-conjugated network. The associated band structure of the prototypical material \ce{PC6} exhibits a direct band gap at the M point and Dirac points above and below the Fermi level at the K point. Isostructural substitutions leave the general shape of the band structure unchanged. Two strategies were presented to shift the Fermi level: in order to access those Dirac points, substitution with other heteroatoms and addition of hydrogen. By substituting phosphorous with, \textit{e.g.}, germanium (two valence electron less per unit cell than \ce{PC6}) the Dirac point below, and by substition with, \textit{e.g.}, sulfur (two valence electron more per unit cell than \ce{PC6}), the Dirac point above the Fermi level can be accessed. Heteroatoms can also be used as binding sites for hydrogen, which adds more electrons to the system. Thus, the Dirac point above the initial Fermi level can be accessed, as demonstrated for \ce{HPC6} and \ce{HAsC6}. Furthermore, only one heteroatom per unit cell can be substituted isoelectronically. Here, the Dirac points are gapped due to the broken inversion symmetry and systems with small band gap ($E_\mathrm{g} < 0.2\,\si{\electronvolt}$) and effective masses ($\nicefrac{m^*}{m_\mathrm{e}} = 0.02$) are obtained. A search for non-trivial topology was conducted. However, no system exhibits a non-zero $Z_2$ invariant.

\begin{acknowledgments}
Financial support by Deutsche Forschungsgemeinschaft (CRC 1415) is acknowledged. The authors thank ZIH Dresden for the use of computational resources.
\end{acknowledgments}

\section*{Supporting Information}
\begin{center}
    \includegraphics[width=0.9\textwidth]{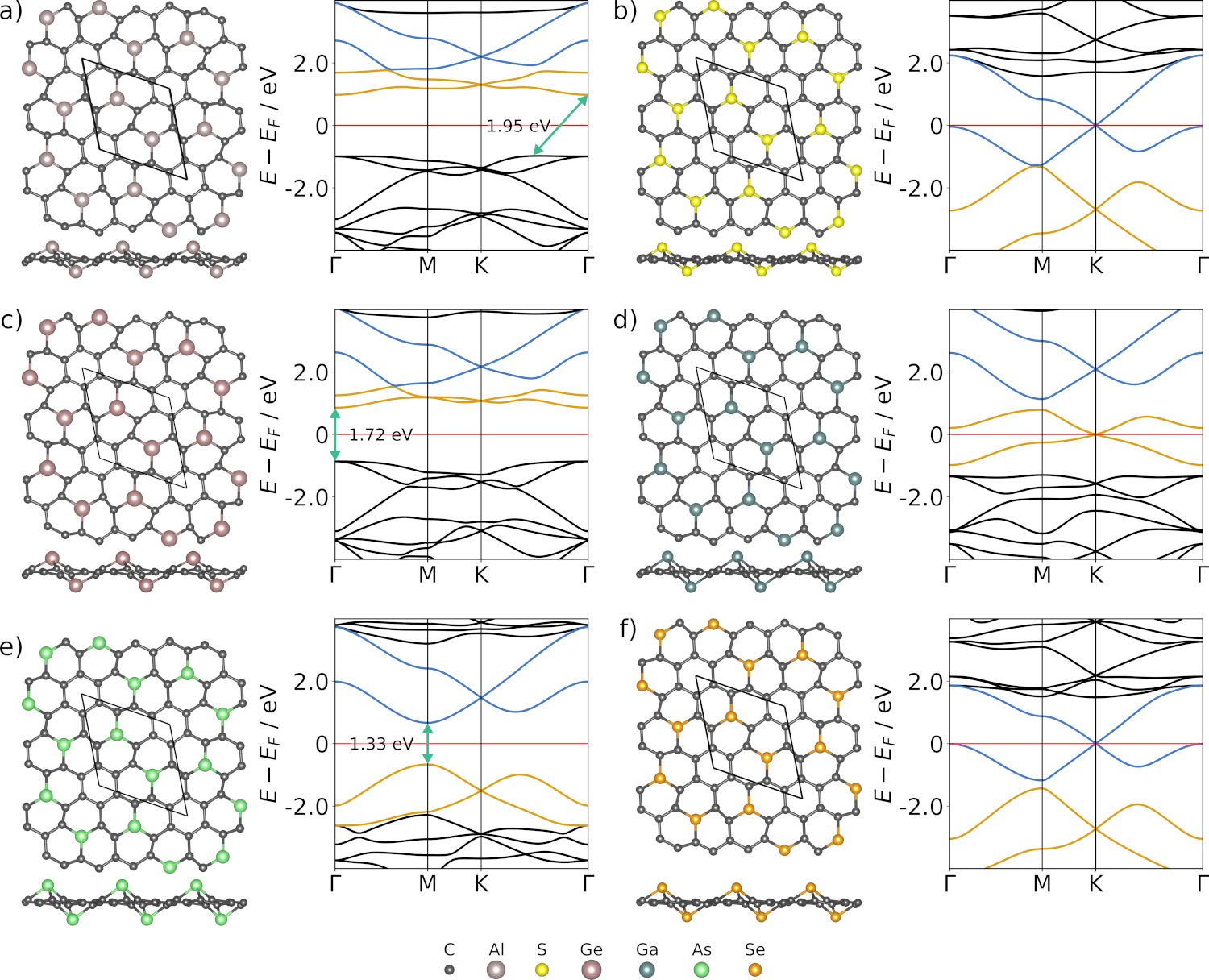}
    \captionof{figure}{Structures and band structures for (a) \ce{AlC6}, (b) \ce{SC6}, (c) \ce{GaC6}, (d) \ce{GeC6}, (e) \ce{AsC6}, and (f) \ce{SeC6}. Characteristic bands are colored according to Fig.~\ref{fig:PC6_fig1}c.}\label{fig:SI_PC6_plains}
\end{center}
\newpage
\begin{center}
    \captionof{table}{Effective hole ($m^*_\mathrm{h}$) and electron ($m^*_\mathrm{e}$) masses in units of the electron rest mass $m_\mathrm{e}$ for \ce{AlC6}, \ce{PC6}, \ce{GaC6}, and \ce{AsC6}. $\Lambda$ is a point between $\Gamma$ and $\mathrm{K}$.}\label{tab:SI_PC6_plain_effective}
    \begin{tabular}{|c|c|c|}\hline
    	& $\nicefrac{m^*_\mathrm{h}}{m_\mathrm{e}}$ &	$\nicefrac{m^*_\mathrm{e}}{m_\mathrm{e}}$	\\\hline
\ce{AlC6} & \thead{2.5 ($\Lambda \rightarrow \mathrm{K}$)\\2.8 ($\Lambda \rightarrow \Gamma$)} &1.1\\\hline
\ce{PC6} & \thead{0.5 ($\Gamma \rightarrow \mathrm{M}$)\\0.2 ($M \rightarrow \mathrm{K}$)} & \thead{0.3 ($\Gamma \rightarrow \mathrm{M}$)\\0.2 ($M \rightarrow \mathrm{K}$)}\\\hline
\ce{GaC6}	&	6.7	&	1.0\\\hline
\ce{AsC6} & \thead{0.5 ($\Gamma \rightarrow \mathrm{M}$)\\0.2 ($M \rightarrow \mathrm{K}$)} & \thead{0.4 ($\Gamma \rightarrow \mathrm{M}$)\\0.3 ($M \rightarrow \mathrm{K}$)}\\\hline
    \end{tabular}
\end{center}

\begin{center}
    \captionof{table}{Hirshfeld charges for \ce{PC6}-type systems with one type of heteroatom. $h(X)$ is the charge of the heteroatom, $h(C_X)$ is the charge of the carbon atom next to the heteroatom (kagome sublattice), and $h_S$ is the charge of carbon atoms in the six-membered ring (hexagonal sublattice).}\label{tab:SI_PC6_plain_charges}
    \begin{tabular}{|c|c|c|c|}\hline
     & $h(X)$ & $h(C_X)$ & $h(C_S)$\\\hline
    \ce{NC6} & -0.008 & 0.026 & -0.023\\\hline
    \ce{AlC6} & 0.445 & -0.135 & -0.013\\\hline
    \ce{SiC6} & 0.335 & -0.118 & -0.001\\\hline
    \ce{PC6} & 0.185 & -0.059 & -0.003\\\hline
    \ce{SC6} & 0.340 & -0.067 & -0.046\\\hline
    \ce{GaC6} & 0.368 & -0.110 & -0.013\\\hline
    \ce{GeC6} & 0.205 & -0.073 & 0.005\\\hline
    \ce{AsC6} & 0.246 & -0.071 & -0.011\\\hline
    \ce{SeC6} & 0.416 & -0.084 & -0.055\\\hline
    \end{tabular}
\end{center}
\vspace{5cm}
\newpage

\begin{center}
    \includegraphics[width=0.7\textwidth]{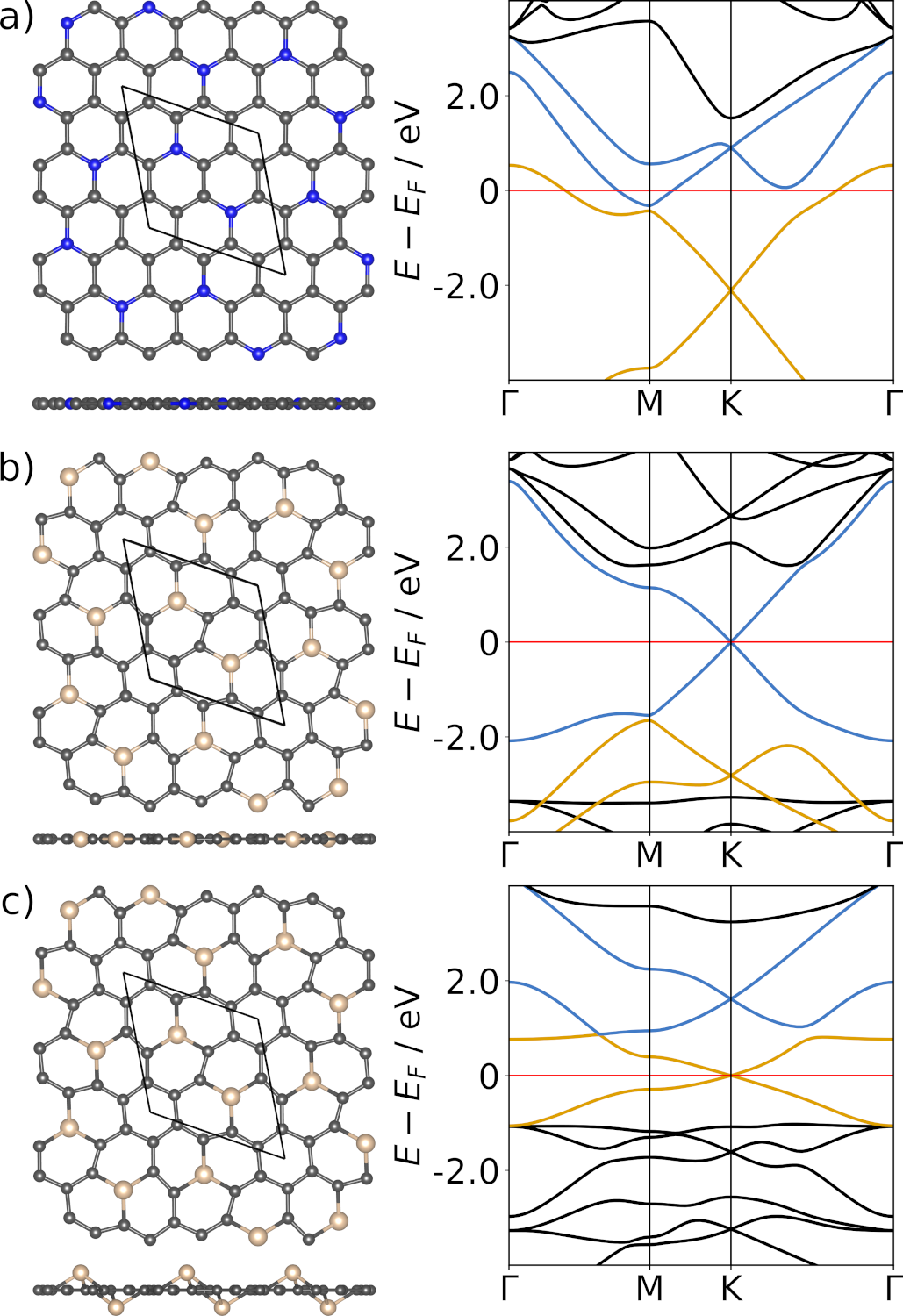}
    \captionof{figure}{Structures and band structures of flat \ce{PC6}-type materials (a) \ce{NC6}, (b) \ce{SiC6}, and (c) artificially puckered \ce{SiC6}. \ce{NC6} is a metallic system, where the characteristic bands overlap energetically. In the band structure of the flat \ce{SiC6}, the Fermi level cuts the upper Dirac cone (blue bands), instead of the expected lower Dirac point corresponding to the band structure of \ce{GeC6} (Fig.~\ref{fig:SI_PC6_plains}c). For an artificially puckered \ce{SiC6} structure in panel (c), the Fermi level cuts the lower Dirac cone. Carbon is grey, nitrogen is blue, silicon is beige. Coloring according to Fig.~\ref{fig:PC6_fig1}c.}\label{fig:SI_PC6_flat}
\end{center}

\newpage

\begin{center}
    \includegraphics[width=0.7\textwidth]{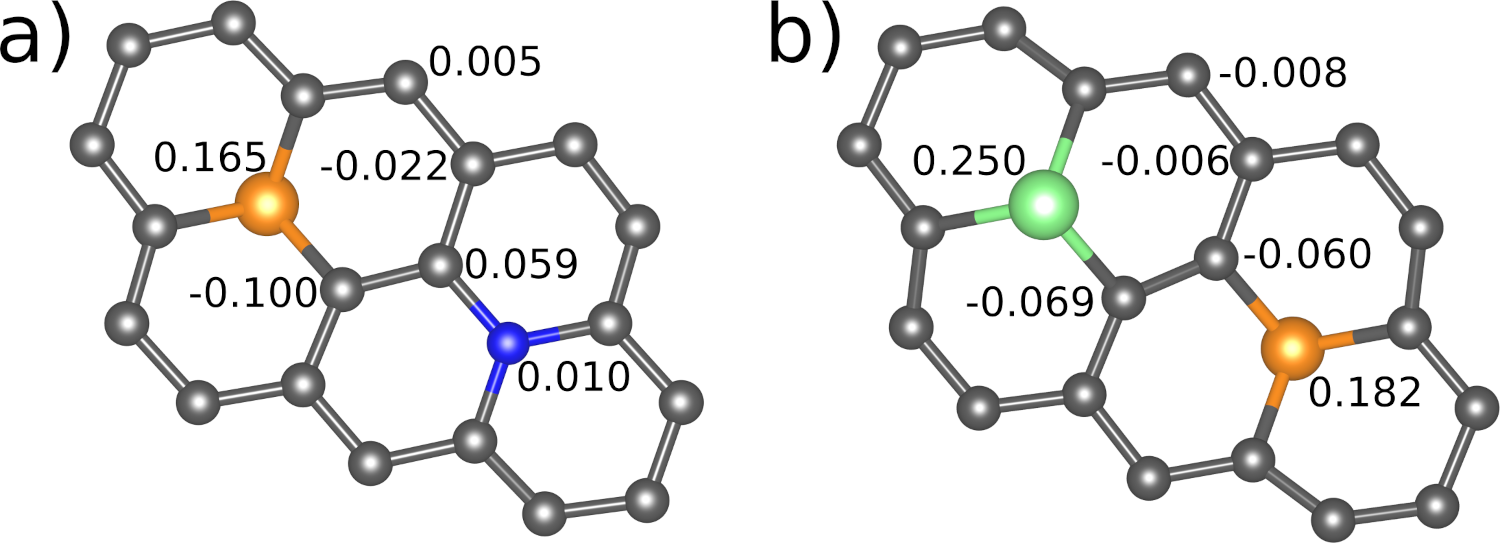}
    \captionof{figure}{Hirshfeld charges of structures with two types of heteratoms per unit cell: (a) \ce{PNC12} and (b) \ce{PAsC12}. Carbon is grey, nitrogen is blue, phosphorous is orange, arsenic is green.}\label{fig:SI_PC6_mixed_charges}
\end{center}

\begin{center}
    \captionof{table}{Effective hole ($m^*_\mathrm{h}$) and electron ($m^*_\mathrm{e}$) masses in units of the electron rest mass $m_\mathrm{e}$ for \ce{PNC12} and \ce{PAsC12}.}\label{tab:SI_PC6_mixed_electronics}
    \begin{tabular}{|c|c|c|}\hline
    	& $\nicefrac{m^*_\mathrm{h}}{m_\mathrm{e}}$ &	$\nicefrac{m^*_\mathrm{e}}{m_\mathrm{e}}$	\\\hline
\ce{PNC12}	&	\thead{1.2 ($\Gamma \rightarrow \mathrm{M}$)\\1.8 ($M \rightarrow \mathrm{K}$)}	&	\thead{0.6 ($\Gamma \rightarrow \mathrm{M}$)\\0.4 ($M \rightarrow \mathrm{K}$)}\\\hline
\ce{PAsC12} & \thead{0.5 ($\Gamma \rightarrow \mathrm{M}$)\\0.3 ($M \rightarrow \mathrm{K}$)} & 0.3\\\hline
    \end{tabular}
\end{center}

\vspace{1.5cm}

\begin{center}
    \includegraphics[width=0.7\textwidth]{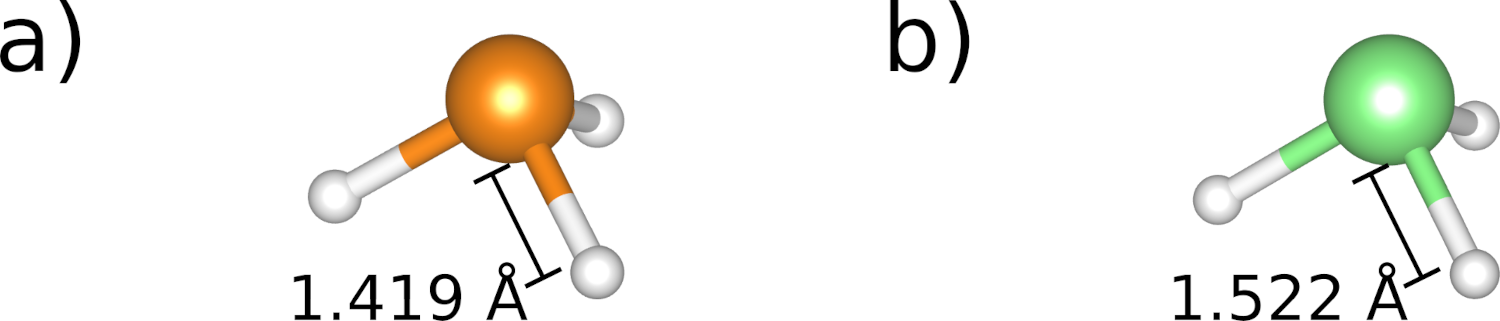}
    \captionof{figure}{Bond lengths in (a) phosphine and (b) arsenine. Geometries were calculated using B3LYP/TZ2P as implemented in AMS/ADF with scalar ZORA relativistics for arsenine~\cite{stephens_ab_1994,van_lenthe_optimized_2003,te_velde_chemistry_2001,van_lenthe_relativistic_1993,van_lenthe_relativistic_1994,van_lenthe_geometry_1999}. Hydrogen is white, phosphorous is orange, and arsenic is green.}\label{fig:SI_PC6_phosphine-arsenine}
\end{center}
\newpage
\begin{center}
    \includegraphics[width=0.7\textwidth]{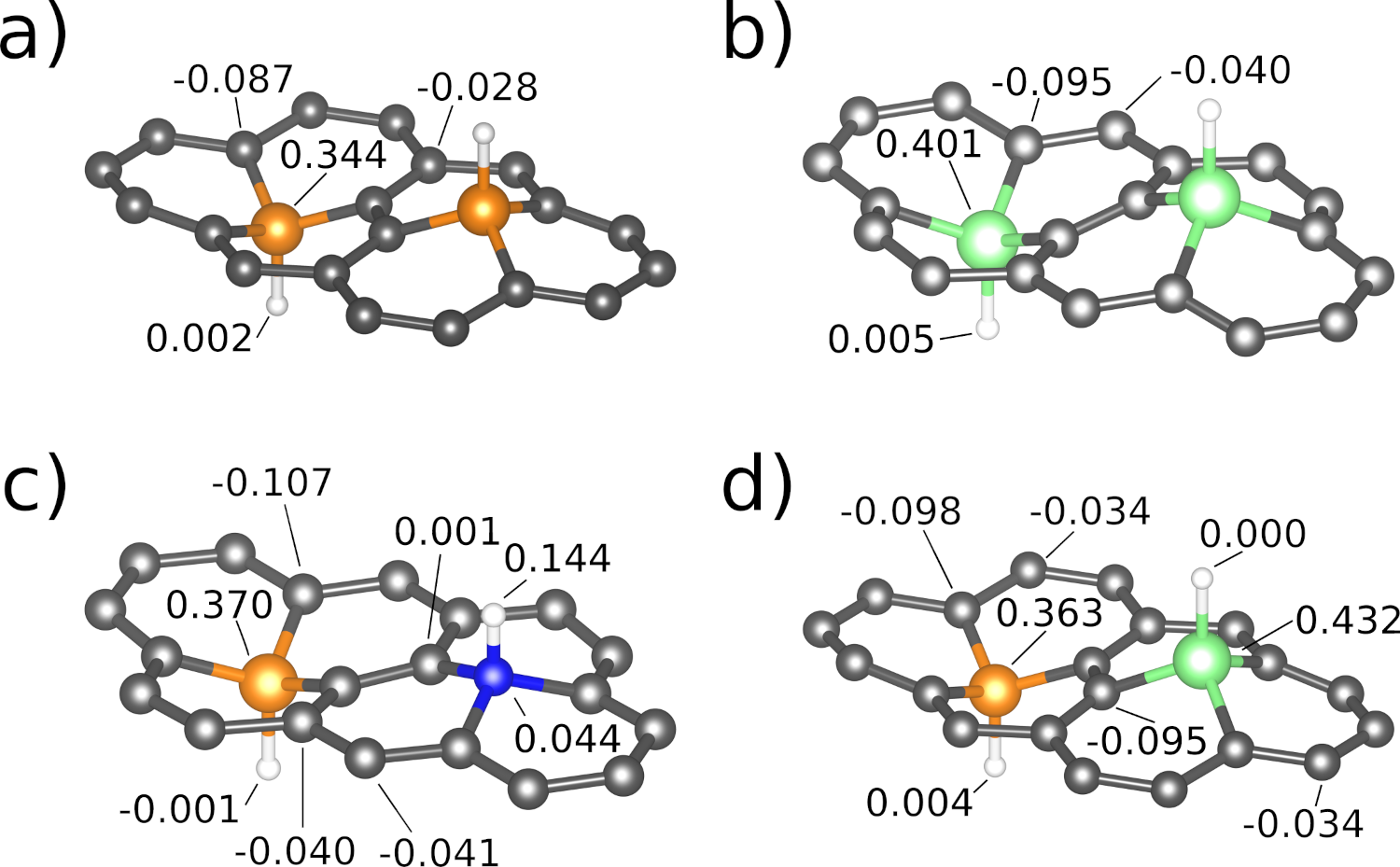}
    \captionof{figure}{Hirshfeld charges structures with hydrogen adatoms: (a) \ce{HPC6}, (b) \ce{HAsC6}, (c) \ce{H2PNC12}, and (d) \ce{H2PAsC12}. Hydrogen is white, carbon is grey, nitrogen is blue, phosphorous is orange, arsenic is green.}\label{fig:SI_PC6_hydrogenated_charges}
\end{center}

\begin{center}
    \captionof{table}{Effective hole ($m^*_\mathrm{h}$) and electron ($m^*_\mathrm{e}$) masses in units of the electron rest mass $m_\mathrm{e}$ for \ce{H2PNC12} and \ce{H2PAsC12}.}\label{tab:SI_PC6_hyd_mixed_effective}
    \begin{tabular}{|c|c|c|}\hline
    	& $\nicefrac{m^*_\mathrm{h}}{m_\mathrm{e}}$ &	$\nicefrac{m^*_\mathrm{e}}{m_\mathrm{e}}$	\\\hline
\ce{H2PNC12}	&	\thead{0.3 ($\Gamma \rightarrow \mathrm{M}$)\\0.4 ($M \rightarrow \mathrm{K}$)}	&	0.4\\\hline
\ce{H2PAsC12} & 0.2 & \thead{0.2 ($\Gamma \rightarrow \mathrm{M}$)\\0.3 ($M \rightarrow \mathrm{K}$)}\\\hline
    \end{tabular}
\end{center}

\end{document}